\documentclass{jaa}
\usepackage{natbib}
\usepackage{graphicx}

\usepackage{authblk}

\begin{document}\sloppy

\title{  Model Independent Probe of Variation of Cosmic Opacity with Redshift }

\author{Savita Gahlaut\textsuperscript{1} and Meetu Luthra\textsuperscript{2}}

\affilOne{\textsuperscript{1}Department of Physics, Deen Dayal Upadhyaya College, (University of Delhi),\\ New Delhi-110078, India\\}

\affilTwo{\textsuperscript{2} Department of Physics, Bhaskaracharya College of Applied Sciences, (University of Delhi),\\ New Delhi-110075, India}

\twocolumn[{

\maketitle 

\corres{savitagahlaut@ddu.du.ac.in}

\msinfo{2026}{2026}

\begin{abstract}
 
Cosmic opacity may vary spatially due to the inhomogeneous distribution of dust, its grain properties, and the efficiency of photon attenuation. 
In this work, we present a model independent method to investigate the variation of  cosmic opacity with redshift. Using strong gravitational lensing data we construct the opacity independent comoving distance function and we use latest  supernovae type Ia (SNe Ia) Pantheon+ data to estimate the opacity dependent comoving distances. Using the distance duality equation, opacity parameter is constrained. 
 Our analysis indicates a transparent Universe on average over  the  redshift range ($0.01 \leq z \leq 2.26137$) of Pantheon+ sample. However, if we split the dataset into subsamples  with redshift bins of width $\bigtriangleup z = 0.1$, we find appreciable deviation from the transparency in several redshift intervals. Particularly, in the redshift range  $0.3 < z \leq 0.4$, the opacity parameter is $\epsilon = -0.4283^{+0.1914}_{-0.2027}$. The current SNe Ia observations indicate the variation of opacity parameter with redshift. These results may have  a significant impact on  the values of the cosmological parameters deduced from the SNe Ia observations.

\end{abstract}


}]

\doinum{12.3456/s78910-011-012-3}
\artcitid{\#\#\#\#}
\volnum{000}
\year{0000}
\pgrange{1--}
\setcounter{page}{1}
\lp{1}

\section{Introduction}
The cosmologists use a set of variables  that define the evolution of the universe in terms of measurable quantities or parameters.
Various independent sets of observations are used to estimate the values of these parameters. 
 One such parameter being the deceleration parameter $q_o$ and a negative $q_{o}$ means an accelerating Universe at the present epoch.
Type Ia supernovae (SNe Ia) observations indicate an accelerated expansion at present \citep{Riess1, Perl}. Evidence of this acceleration was considered to be SNe Ia appearing fainter than what they would appear in a decelerating universe.
 Gravitational forces, from ordinary matter and energy, alone result in deceleration but introducing a cosmological constant and/or a hypothetical substance, called dark energy, with negative pressure can result in repulsive forces and hence can account for an accelerating Universe.

At the same time many other justifications for this dimming of SNe 1a were suggested. One such justification is photon number not being conserved along the ray path. 
 This violation of conservation of photon flux can be due to some astrophysical effects such as photon attenuation due to interstellar dust, gas or plasmas in the host galaxy, intervening galaxies or the Milky Way \citep{Menard,Bovy}.
Other reasons for this violation could be reappearing or disappearing of photons due to coupling with exotic particles in the presence of extra-galactic magnetic fields \citep{Avgoustidis}. Non-conservation of photon flux as it travels towards the earth from the source can affect the  measurements of luminosity distance using SNe Ia observations and in consequence the estimated value of the cosmological parameters. It can also distort the observed evolution of the global stellar mass density with redshift. Thermal radiation of the intergalactic dust can add to the cosmic microwave background radiation (CMBR) and hence can alter the estimation of the cosmological parameters deduced from the observations \citep{Vav,Narlikar}.

The evidence of accelerating universe also comes from the observations of large scale structures, baryon acoustic oscillations and cosmic microwave background anisotropy .
However, the question remains whether the accelerated expansion of the Universe alone is causing the dimming or opacity of the Universe can atleast partially account for the reduction of SNe Ia luminosity. 

The  opacity of the universe has been examined using distance duality relation in multiple works with diverse cosmological observations. Distance duality relation expressed as:$D_{L} = D_{A}(1+z)^{2}$; relates the luminosity distance $D_{L}$ and the angular diameter distance $D_{A}$ \citep{Eth}. The relation is true for any cosmological model if the following two properties are satisfied : (a) It is described by Riemannian geometry (b)The photon number is conserved and photons travel along the null geodesics \citep{Ellis}. A model independent approach is used in some works while some specific cosmological models are used in others. 
\citet{Ma} obtained constraint on cosmic opacity using a combination of the observed and simulated data of strongly lensed SNe Ia and regular SNe Ia. \citet{Jun-Jie} proposed a method to constrain cosmic opacity parameter by combining the gravitational wave (GW) observations with SNe Ia data in corresponding redshift ranges . 
\citet{Zhou} probed cosmic opacity at high redshifts using Gaussian process (GP) by reconstructing the luminosity distance from the GW data and comparing it with that deduced from the Pantheon SNe Ia and gamma ray burst (GRB) data in flat $\Lambda$CDM model.
 \citet{Liu} considered the nonlinear relation between X-ray and UV emissions of quasars to derive the luminosity distances and Hubble parameter H(z) measurements along with simulated GW events were used to derive the opacity-independent distances.
While \citet{Bing Xu} probed the opacity using 
the  Pantheon SNe Ia and the  Hubble parameter (H(z)) data in $\Lambda$CDM and wCDM models with or without spatial curvature, \citet{Fu}  used  the combination of  future gravitational wave  and SNe Ia observations to measure cosmic opacity.  All of these works report a transparent Universe or small opacity on an average for the redshift range of the dataset used within $1\sigma$ error range; see Table 2.  
 \citet{Eli} used Bayesian machine learning to probe the cosmic opacity and \citet{Ying Yang}  measured  curvature parameter and cosmic opacity by combining the  observations of HII galaxies acting as standard candles with the H(z) observations and reported a slightly opaque Universe.

In most of the works, the average cosmic opacity is probed for the entire redshift range of the datasets used. However, it is an integral quantity that depends upon the dust distribution along the path from source to observer, proper dust density and dust extinction efficiency. 
It is a spatially dependent quantity and varies with the frequency and redshift
\citep{Aguirre1, Aguirre2, Aguirre3, Corasaniti, Vavrycuk1,Vavrycuk2}. 
In this paper we present a model independent method to
examine the variation of the cosmic opacity with redshift. We use strong gravitational lensing data to estimate the opacity independent angular diameter distances and latest SNe 1a Pantheon+ data to measure luminosity distances at the corresponding redshifts. In an opaque universe with optical depth $\tau(z)$, the flux received from SNe will be diminished by a factor of $e^{-\tau(z)}$. The optical depth varies with redshift $z$ and is zero for a transparent Universe. The luminosity distance $D_{L_{SN}} $ measured from SNe 1a data is related to the actual luminosity distance $D_{L_{0}} $ as: $  D_{L_{0}}(z) =  D_{L_{SN}}(z) e^{-\tau(z)/2}$. Using this relation, $\tau(z)$ is estimated in different redshift bins. The methodology and data sets used in the analysis are described in section 2. In Section 3, model independent constraints on cosmic opacity at various redshifts are reported. Conclusions and discussions are presented in section 4.

\section{Methodology and Data}

In a homogeneous and isotropic Universe, represented by the Friedmann-Lema$\hat{\textsf{i}}$tre-Robertson-Walker (FLRW) metric, the angular diameter distance $D_{A}(z)$ and the luminosity distance $D_{L}(z)$ of a source at redshift $z$  are:
\begin{equation}
D_{A}(z) = \frac{c}{H_{0}(1+z)} d(z) 
\end{equation}
\begin{equation}
D_{L}(z) = (1+z)\frac{c}{H_{0}} d(z) 
\end{equation} 
where c is the speed of light and $d(z)$ is the dimensionless transverse comoving distance. In  FRLW cosmology $d(z)$ is expressed as :
\begin{equation}
d(z) = \left\lbrace
\begin{array}{cc} D_{c},&\,\, K=0\\ &\\
\frac{1}{\sqrt{|\Omega_{K}}|}\sinh{(\sqrt{|\Omega_{K}|}D_{c})},&\, K=-1\\ &\\ \frac{1}{\sqrt{|\Omega_{K}|}}\sin{(\sqrt{|\Omega_{K}|} D_{c})},&K=1

\end{array}\right.
\end{equation}

with  $D_{c} = \int_{0}^{z}\frac{H_{0}}{H(z')}dz'$ and $\Omega_{K} = -K c^{2}/{a^{2}_{0}H^{2}_{0}}$. Here $H(z)$ is the Hubble parameter and $H_{0}$, $a_{0}$ are the present values of Hubble parameter and scale factor respectively.

 In a transparent Universe, the comoving distance $d_{L}(z)$ measured from the luminosity distance is related to the comoving distance $d_{A}(z)$ deduced from the angular diameter distance as:
\begin{equation}
d_{L}(z) = (1+z)^{2} d_{A}(z)
\end{equation}
However in an opaque Universe,  the flux received from a source (SNe Ia) will be diminished and the actual luminosity distance $d_{L_{0}}$ should be smaller than the observed one. This effect can be characterized  by a factor $e^{-\tau(z)}$, where $\tau(z)$ is the optical depth of the Universe at redshift $z$ due to cosmic absorption. The true luminosity distance of an SNe Ia is: 
\begin{equation}
d^{2}_{L_{0}} = d^{2}_{L_{Ob}} e^{-\tau(z)}
\end{equation}
 On the other hand, the angular diameter distance $d_{A}$ is not affected by the opacity of the Universe.
 
To use equation (4), for the study of cosmic opacity, one needs to estimate the opacity independent distance $d_{A}(z)$ and the opacity dependent distance $d_{L}(z)$ from observations at the same redshift. To match the angular diameter distances deduced from SGL data with the luminosity distances deduced from SNe Ia observations at the same redshifts, as discussed in our previous work \citep{SG}, we determine $d_{A}(z)$ in a model independent way by fitting a third order polynomial to the strong gravitational (SGL) data. The polynomial is expressed as:
 \begin{equation}
 d(z) = z + a_{1} z^{2} + a_{2}z^{3}
\end{equation}   
where $a_{1}$ and $a_{2}$ are the free parameters to be optimized from the observations. The polynomial is selected such that it satisfies the initial conditions $d(0) = 0$ and $d'(z) = 1$ \citep{Wei20,Qi21}. Further, it is found that a third order polynomial is good enough to fit the SGL data and the fitness results are not improved by taking
higher order polynomials. Increasing the order of polynomial means increasing the number of free parameters and hence no improvement of fitness results.

\subsection{Strong Gravitational Lensing Data}

Multiple images of a background source are observed due to bending of light by the strong gravity of the massive objects (galaxies or cluster of galaxies) present along the path of light. If the source and the massive object (called the lens) are perfectly aligned along the line of sight, a bright circular  ring called the Einstein ring is formed. 
The radius of the Einstein ring,  $\theta_{E}$, depends on the mass of the lensing object and the angular diameter distances between the source, lens and observer. For the singular isothermal sphere (SIS) mass distribution model, generally used for the early type galaxies, the Einstein radius is:
\begin{equation}
\theta_{E} = 4\pi \frac{\sigma_{SIS}^{2}}{c^{2}} \frac{D_{ls}(z)}{D_{s}(z)}
\end{equation}
  where $\sigma_{SIS}$ is the velocity dispersion of stars in the lens and $D_{ls}(z)$, $D_{s}(z)$  are the angular diameter distances from source to lens and from source to observer  respectively.
  Strong gravitational lensing data comprising of 161 early type lens galaxies, compiled by \citet{Chen19}, furnishes the observed values of $\theta_{E}$ and $\sigma_{ap}$ (the luminosity averaged central velocity dispersion measured within the aperture) along with the uncertainties. The SGL systems in the sample are selected from the LSD, SL2S, SLACS, S4TM, BELLS and BELLS GALLERY surveys. To ensure that the lens galaxies in the sample are spherically symmetric, only early-type lens galaxies with E or S0 morphologies which do not have any massive neighbours or significant substructure are selected. Since the lenses in the sample are chosen from different surveys, the spectroscopically measured $\sigma_{ap}$ is normalised to the velocity dispersion within a circular aperture of radius $R_{eff}/2$, where $R_{eff}$ represents the half-light radius of the lens \citep{Jor,Jor1,Cap}. The normalised velocity dispersion $\sigma_{0}$ is expressed as:
\begin{equation}
\sigma_{0} = \sigma_{ap}(\theta_{eff}/(2\theta_{ap}))^{\nu}
\end{equation} 
where $\theta_{eff} = R_{eff}/D_{l}$ and $\nu = -0.04$ is the correction factor fitted from the samples of observations. Various studies  \citep{Cao1,Koop,Koop1,Treu} of the mass  density profiles in elliptical galaxies in SGL systems, modelled using a general power law mass density profile $\rho(r)$  and luminosity density of stars $v(r)$ ($\rho(r) \propto r^{-\alpha} $ and $v(r) \propto r^{-\delta} $, where $\alpha$ and $\delta$ are free parameters and $r$ is the radial coordinate from the center of the lens),
have established that the intermediate-mass, early type elliptical galaxies show the best agreement with the SIS mass distribution model with $\alpha \approx 2$. Therefore, we select only those systems from the catalog whose velocity dispersions lie in the range $200 km s^{-1}\leq \sigma_{ap} \leq 300 km s^{-1}$. Further    
 the velocity dispersion in the SIS model, $\sigma_{SIS}$, may differ from the central velocity dispersion $\sigma_{0}$, Kochanek \citep{Koch} introduced a phenomenological free parameter $f_{E}$ such that $\sigma_{SIS} = f_{E}\sigma_{0}$. The parameter $f_{E}$ accounts for contributions from dark matter halos in velocity dispersion measurements, systematic uncertainties in the measurement of image separation and potential effects of background matter along the line of sight. These factors can affect the image separation by up to $20\%$ constraining $f_{E}$ within the range $\sqrt{0.8} < f_{E} < \sqrt{1.2}$ (\citealt{Cao,Ofek}). Since the primary goal of this analysis is to constrain cosmic opacity and
 we  select only those sources from the sample that are consistent with a SIS model (for which $f_{E} \sim 1$ ), we adopt a single parameter $f_{E}$ for all systems, treat it as a free-parameter, and determine its best fit value along with its probability  function. The best fit values of the relevant cosmological parameters are then obtained by treating $f_{E}$ as a \textit{nuisance} parameter.

  The distance ratio $D^{ob}(z_{l},z_{s}) \equiv \frac{D_{ls}}{D_{s}} =\frac{d_{ls}}{d_{s}}$ is evaluated and the free parameters $a_{1}$ and $a_{2}$ in equation (6) are optimized. The likelihood function to be maximized, to find the optimized values of parameters, for the SGL data is:
\begin{equation}
L_{1} = e^{-\chi_{1}^{2}/2}
\end{equation}
where:
\begin{equation}
\chi_{1}^{2} = \sum_{i=1}^{N} \left(\frac{D^{th}(z_{l,i},z_{s,i},a_{1},a_{2})-D^{ob}(\sigma_{0,i},\theta_{E,i})}{D^{ob}\Delta D^{ob}_{i}}\right)^{2}
\end{equation}
here $N = 102$ is the number of data points used and $\Delta D^{ob}$ is the uncertainty in the value of $D^{ob}$ given by:
\begin{equation}
\Delta D^{ob} = \sqrt{\left(\frac{\Delta\theta_{E}}{\theta_{E}}\right)^{2}+ 4\left(\frac{\Delta\sigma_{0}}{\sigma_{0}}\right)^{2}}
\end{equation}
where $\Delta\theta_{E}$ and $\Delta\sigma_{0}$ are the uncertainties in the Einstein radius and velocity dispersion measurements respectively.
For detailed discussion of the SGL data set, selection criteria and systematics see \citet{SG}.

To obtain one-dimensional likelihood functions and two-dimensional confidence contours,
the marginalization is carried out by computing the integral of the form: 
 \begin{equation}
 L(p_{1}) = \int L(p_{1},p_{2}) P(p_{2}) dp_{2}
\end{equation}    
 where $p_{2}$ is the nuisance parameter to be marginalized and $P(p_{2})$ is the uniform prior for $p_{2}$ such that: $P(p_{2}) = 1$ if $p_{2,min}< p_{2}< p_{2,max}$ else $P(p_{2}) = 0$ (see Table 1 for parameter ranges).

\subsection{Pantheon+ SNe Ia Data}
\citet{Scolnic} recently published the Pantheon+ SNe Ia data set wherein the corrected apparent magnitudes ($m_{corr} = \mu + M_{B}$), for 1701 supernova light curves from 1550 distinct type Ia supernovae, are reported. The redshift range of the data is from $z = 0.001$ to $2.26$ and the distance moduli are fitted using a SALT2 model. The data uncertainties, including both  statistical and systematic, are represented by a $1701$x$1701$ covariance matrix $C$ \citep{Brout,Riess}.

 The distance modulus $\mu$ of a supernova, with apparent magnitude $m$ and absolute magnitude $M'$, is related to the luminosity distance $D_{L}(z)$ as:
\begin{equation}
\mu^{th}(z) = m - M' = 5 log(D_{L}/Mpc)+ 25
\end{equation}
For an opaque Universe, with the optical depth $\tau(z)$, the expression for $\mu^{th}(z)$ is modified as:

\begin{equation}
\mu^{th}(z) = m - M' = 5 log(\frac{c}{H_{0}} (1+z) e^{\tau(z)/2} d_{L}(z)/Mpc)+ 25
\end{equation}
We take optical depth $\tau(z)$  as a parametrized function of the redshift:

\begin{equation}
\tau(z) = 2\epsilon z
\end{equation}
which is independent of frequency in the optical band region \citep{Liao1,Liao2}. For a transparent Universe $\epsilon = 0$.

The likelihood function for the SNe Ia data is $L_{2} = e^{-\chi_{2}^{2}/2}$  with:
\begin{equation}
\chi_{2}^{2} =  \mathbf{{\bigtriangleup\mu}^{\dagger} \cdot C^{-1} \cdot \mathbf{\bigtriangleup\mu}}
\end{equation}
where the vector $\bigtriangleup\mu_{i} \equiv \mu^{th}(z_{i},\textbf{p}) -  \mu^{ob}(z_{i})$ is formed from the residuals of the data and $\textbf{p} = (\epsilon,a_{1},a_{2},M)$ is the vector formed by the parameters to be fitted from the sample.
Owing to the degeneracy exhibited by  the factor $5 log_{10}(c/ H_{0}Mpc)+25$ and absolute magnitude $M'$, the two are jointly fitted as the parameter $M =  M' +5 log_{10}(c/ H_{0}Mpc)+25$. The SH0ES Cepheid host-galaxy distances provided with the data,  used to break degeneracy between parameters $M'$ and $H_{0}$, are not considered in the analysis. Further, the low redshift ($z \leq 0.01$) samples are also excluded to avoid strong peculiar velocity dependence of the sample.  

The joint log-likelihood function for SGL and SNe datasets is:
\begin{equation}
\textit{ln}(L_{tot}) = -0.5(\chi_{1}^{2} + \chi_{2}^{2})
\end{equation}
 By maximizing the likelihood function $L_{tot}$, the best fitting values of the parameters are obtained. 
First we extremize the likelihood function taking all the 1590 data points with $z > 0.01$ in the Pantheon+ sample and determine the best fit values of the parameters. Next we divide the Pantheon+ sample into subsamples with redshift range $\bigtriangleup z = 0.1$ and crop the Pantheon+ covariance matrix $C$ to isolate the $n$ x $n$ dimensional covariance matrix corresponding to the $n$ data points in a subsample. Markov Chain Monte Carlo (MCMC) analysis is performed, using Python module \textit{emcee} \citep{emcee},  to estimate the best fit values and $1\sigma$ confidence intervals for each subsample. The uniform priors of the parameters used in the analysis are chosen large enough so that they do not impact the best fit values and are presented in Table 1 \label{Priors}.

\vspace{-2em}
\begin{table}[htb] 
\tabularfont
\caption{The prior distributions of parameters}

\begin{tabular}{|c|c| }
\topline
\textbf{Parameter}  & \textbf{Priors} \\
\hline
&\\
$\epsilon$ & Uniform ($-1 , 1)$ )\\
&\\
$a_{1}$ & Uniform ($-0.5 , 0.1)$ )\\
&\\
$a_{2}$ & Uniform ($-0.001 , 0.1)$ )\\
&\\
 M & Uniform ($20 , 30)$ )\\
&\\
$f_{E}$ & Uniform ($0.8 , 1.2)$ )\\
&\\
\hline
\end{tabular} \\
\label{tab: Priors}
\end{table}

\section{Results} 

We start with  the entire cosmology-only Pantheon+ sample  and  
the analysis shows that the Universe is transparent on average within $68.3\%$ confidence level  over the redshift range ($0.01 < z \leq 2.26137$) of the sample. 
The optimized value of the opacity parameter, $\epsilon = 0.0095^{+0.0510}_{-0.0878} $, is in agreement with the results quoted in previous works ,for e.g., \citet{Chen, Ma,Jun-Jie,Zhou,Liu,Bing Xu,Fu} etc. For comparison, the results of the above mentioned works are presented in Table 2.
The best fit value of parameter $M = 23.8124$; corresponds to $H_{0} = 73.3$km/s/Mpc, if we take
 $M' = -19.25\pm 0.03$ (obtained using the SH0ES Cepheid host-galaxy distances; \citet{Peri}), which is in good  agreement with \citet{Brout}.

\begin{table*}[htb] 
\tabularfont
\begin{center}
\caption{Best fit values of opacity parameter reported in earlier works}

\begin{tabular}{|c|c|c| }
\topline
\textbf{Data Sets}  & \textbf{$\epsilon $} & \textbf{Ref} \\
\hline
&&\\
H(z)+ SNeIa(Union2.1)& $0.0097^{+0.0262}_{-0.0262}$ & \citet{Chen1}\\
&&\\
SLSNeIa(LSST)+ SNeIa(Panth)& $< 0.02$ & \citet{Ma}\\
&&\\
GW+SNeIa(Panth)& $0.04^{+0.026}_{-0.026}$ & \citet{Jun-Jie}\\
&&\\
GW+SNeIa(Panth)& $0.005^{+0.005}_{-0.006}$ & \citet{Zhou}\\
&&\\
QSO+H(z) & $0.086^{+0.072}_{-0.075}$ & \citet{Liu}\\
&&\\

SNeIa(Panth)+H(z) & $0.005^{+0.033}_{-0.033}$ & \citet{Bing Xu}\\
&&\\
GW+WFiRST& $-0.0011^{-0.012}_{-0.012}$ & \citet{Fu}\\
&&\\
SGL+SNeIa(Panth+) &$0.0095^{+0.0510}_{-0.0878} $ & This work\\
&&\\

\hline
\end{tabular} \\
\end{center}

\label{tab: Result Comparison}
\end{table*}

Thereafter we estimate the opacity parameter in different redshift bins. The best fit values of the optimized parameters with $1\sigma$ error in the redshift bins of width $\bigtriangleup z = 0.1$ are presented in Table 3.  In the range $0.01 < z \leq 0.3$ and  for $z  > 0.6$ the data suggests  a transparent Universe within $1\sigma$ error interval. 
Some deviations from a transparent Universe are  detected in the redshift range $0.3 < z \leq 0.5$; see Figure 1. The largest deviation is in the redshift bin  $0.3 < z \leq 0.4$ (186 data points) with $\epsilon = -0.4283^{+0.1914}_{-0.2027} $. In the higher redshift bins the subsample size is smaller making them prone to statistical fluctuations and the
 constraints become relatively weak with larger $1\sigma$ uncertainties.
\citet{Chen} and \citet{Chen1} also reported deviations from cosmic transparency in some redshift regions using H(z), baryon acoustic oscillation (BAO) and Union2 SNeIa data (See Table 3 and 4 in \citet{Chen1}). 

It is also observed that the best fit value of parameter $M =  M' +5 log_{10}(c/ H_{0}Mpc)+25$ increases as the value of $\epsilon$ decreases across the redshift bins, indicating that   the parameter $M$ is sensitive to the value of cosmic opacity parameter. Some recent studies have also reported similar results. For instance, \citet{Mukherjee} investigate a possible late time transition of the absolute magnitude ($M'$) of SNeIA in the Pantheon+ sample and  a transition at $z\sim 1$ is observed. Likewise, \citet{Das} examine the Etherington relation (ER), also known as distance duality relation, using a combination of various SNeIa and BAO measurements, finding some deviations from ER with redshift (see figure 2,4) although on average the data shows a remarkable statistical consistency with the relation.  
 A redshift evolution of $M'$ relative to its local value is also evaluated, if the distance duality relation is assumed to be valid, and the maximum change in its value is constrained to be $\frac{dM'}{dz} = 0.11 \pm 0.08$  for the Pantheon+ sample. 
 The increase in the value of $M$ with decreasing $\epsilon$ corresponds to a reduction in the value of $H_{0}$. Hence, a non-zero $\epsilon$ in some patches can affect the constraints on cosmological parameters. 
 However, in higher redshift bins the uncertainties grow due to statistical fluctuations as well as some residual systematics. Therefore, while estimating  cosmological parameters all these factors need to be considered.
 
 The posterior probability density function of opacity parameter marginalized over the other parameters, in different redshift bins, along with 2D confidence contours are presented in figures 2-11.

\section{Discussion and Conclusions}

This paper highlights the fact that combining the Pantheon+ SNe Ia data with the strong gravitational lensing measurements allows us to constraint cosmic opacity at low redshifts.  A model independent mechanism is used to compare the opacity independent distances estimated from the SGL data with the opacity dependent distances obtained from the SNe Ia observations and
measure the cosmic opacity. In our analysis, from the SGL data comprising 161 SGL systems compiled by \citet{Chen19}, we select only the 102 SGL systems that show the best consistencey with the SIS mass distribution model. However, if all 161  systems are included,  a more general single isothermal ellipsoid (SIE) mass distribution model- where the mass distribution is modelled by a power law with free parameters- must be employed. Although the SIE model is more general, it introduces two additional free parameters that must be optimized along with the cosmological parameters; consequently, no significant improvement in the results is obtained. 
 We divide the Pantheon+ sample into subsamples of  redshift range $\bigtriangleup z = 0.1$ and accordingly crop the Pantheon+ covariance matrix $C$ to isolate the $n$ x $n$ dimensional covariance matrix corresponding to the $n$ data points in a subsample. These subsamples are used to study the  variation of opacity parameter with redshift. It is found that on  average the Universe is  transparent within $1\sigma$ confidence interval over the redshift range of the Pantheon+ data.
Some of the fairly recent works also have similar findings at low redshifts. However some opacity is detected in different redshift intervals and the best fit value of the opacity parameter is  not constant with redshift. The current SNe Ia observations suggest a variation of the cosmic opacity with redshift.
 The reason for the patches, at different redshifts, with opacity could be the presence of intergalactic dust, dust in the host galaxies, coupling of photons with exotic particles,  systematic calibration effects or a combination of all. Whatever may be the cause, the opacity in some regions of the Universe can affect the values of the cosmological parameters estimated from the opacity dependent observations. Therefore while constraining cosmological parameters using SNe Ia observations or other opacity dependent probes it is essential to incorporate the possible effects of cosmic opacity.

\begin{table*}[htb] 
\tabularfont
\caption{Best fit values of the optimized parameters with $1\sigma$ errors for various redshift ranges }

\begin{tabular}{|c|c|c|c|c|c|c| }
\topline
\textbf{Red shift range} & \textbf{Data points} & \textbf{$\epsilon $} & $a_{1}$ & $a_{2}$ & M & $f_{E}$\\
\hline
&&&&&&\\
 $0.01 < z \leq 2.26137$ & 1590 &  $0.0095^{+0.051}_{-0.0878} $ & $-0.2891^{+0.0889}_{-0.05}$ & $0.0420^{+0.0153}_{-0.0209}$ & $23.8124^{+0.007}_{-0.0066}$ & $1.0026^{+0.0113}_{-0.0122}$\\
&&&&&& \\
 $0.01 < z \leq 0.1$ & 630 &  $-0.0138^{+0.1978}_{-0.1825} $ & $-0.2858^{+0.0926}_{-0.0905}$ & $0.0431^{+0.025}_{-0.0246}$ & $23.8148^{+0.0167}_{-0.0176}$ & $1.0025^{+0.0128}_{-0.0129}$\\
&&&&&& \\
 $0.1 < z \leq 0.2$ & 207 & $0.0631^{+0.1775}_{-0.1679}$ & $-0.2893^{+0.1004}_{-0.0846}$ & $0.0434^{+0.0248}_{-0.0259}$ & $ 23.7932^{+0.05}_{-0.0512}$ & $1.0021^{+0.0127}_{-0.013}$ \\
 &&&&&&\\
 $0.2 < z \leq 0.3$ & 259 & $  0.0095^{+0.1753}_{-0.1729}$ & $ -0.2921^{+0.0999}_{-0.0819}$ & $  0.0441^{+0.0231}_{-0.0255}$ & $ 23.8025^{+0.0784}_{-0.0795}$ & $1.0025^{+0.0127}_{-0.0126}$ \\
 &&&&&&\\
 $0.3 < z  \leq 0.4$ & 186 & $  -0.4283^{+0.1914}_{-0.2027}$ & $-0.2792^{+0.0948}_{-0.0955}$ & $ 0.0408^{+0.0264}_{-0.0258}$ & $ 24.1411^{+0.1304}_{-0.1287}$ & $ 1.0018^{+0.013}_{-0.0134}$  \\
 &&&&&& \\
 $0.4 < z  \leq 0.5$ & 98 & $ -0.2149^{+0.2482}_{-0.2361}$ & $ -0.2798^{+0.0945}_{-0.0906}$ & $ 0.0414^{+0.0257}_{-0.0249}$ & $24.0184^{+0.2375}_{-0.2320}$ & $1.0016^{+0.0133}_{-0.0128}$\\
 &&&&&&\\
 $0.5 < z \leq 0.6$ & 81 &  $  0.1401^{+0.3048}_{-0.3099}$ & $ -0.2763^{+0.0945}_{-0.0944}$ & $  0.0407^{+0.0261}_{-0.0253}$ & $23.5912^{+0.3646}_{-0.3086}$ & $1.0015^{+0.013}_{-0.0125}$\\
 &&&&&&\\
 $0.6 < z \leq 0.7$ & 54 & $ 0.0184^{+0.3196}_{-0.3343} $ & $-0.2879^{+0.0984}_{-0.0901}$ & $ 0.0440^{+0.0254}_{-0.0267}$ & $23.7966^{+0.4328}_{-0.4526}$ & $1.0033^{+0.0126}_{-0.0122}$\\
 &&&&&&\\
 $0.7 < z  \leq 0.8 $ & 45 & $-0.0688^{+0.4582}_{-0.4206}$ & $ -0.2786^{+0.0941}_{-0.0963}$ & $ 0.0419^{+0.0273}_{-0.0264}$ & $23.8752^{+0.6704}_{-0.7195}$ & $ 1.002^{+0.0134}_{-0.0121}$\\
 &&&&&&\\
 $ z > 0.8$ & 30  & $-0.0102^{+0.0819}_{-0.088}$ & $-0.266^{+0.0857}_{-0.0962}$ & $ 0.0372^{+0.0272}_{-0.0226}$ & $23.8769^{+0.1742}_{-0.1802}$ & $1.0008^{+0.0125}_{-0.0124}$\\
 &&&&&&\\
\hline
\end{tabular} \\

\label{tab: cosmic_opacity parameter}
\end{table*}

\begin{figure}[!t]
\centering

  \includegraphics[width=0.8\columnwidth]{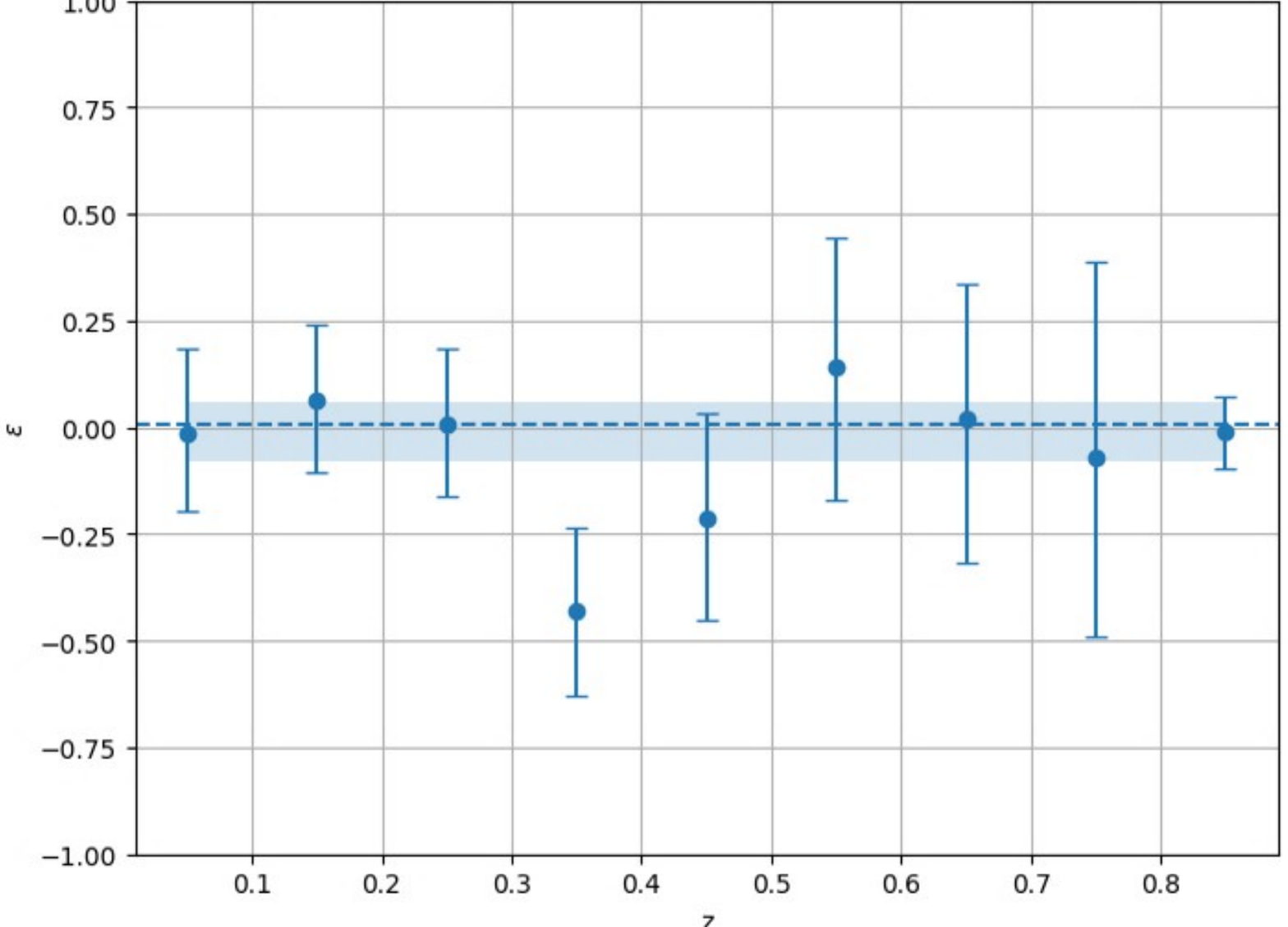}
 \caption{ \small{ Best fit values of cosmic opacity parameter $\epsilon$ with $1\sigma$ uncertainties in different redshift bins. The dashed line represents the best fit value of $\epsilon$ obtained with the complete cosmology-only sample. Shaded region represents the$1\sigma$ uncertainties. }}
  
\end{figure}
 
\begin{figure*}
\centering

  \includegraphics[height=.5\textheight]{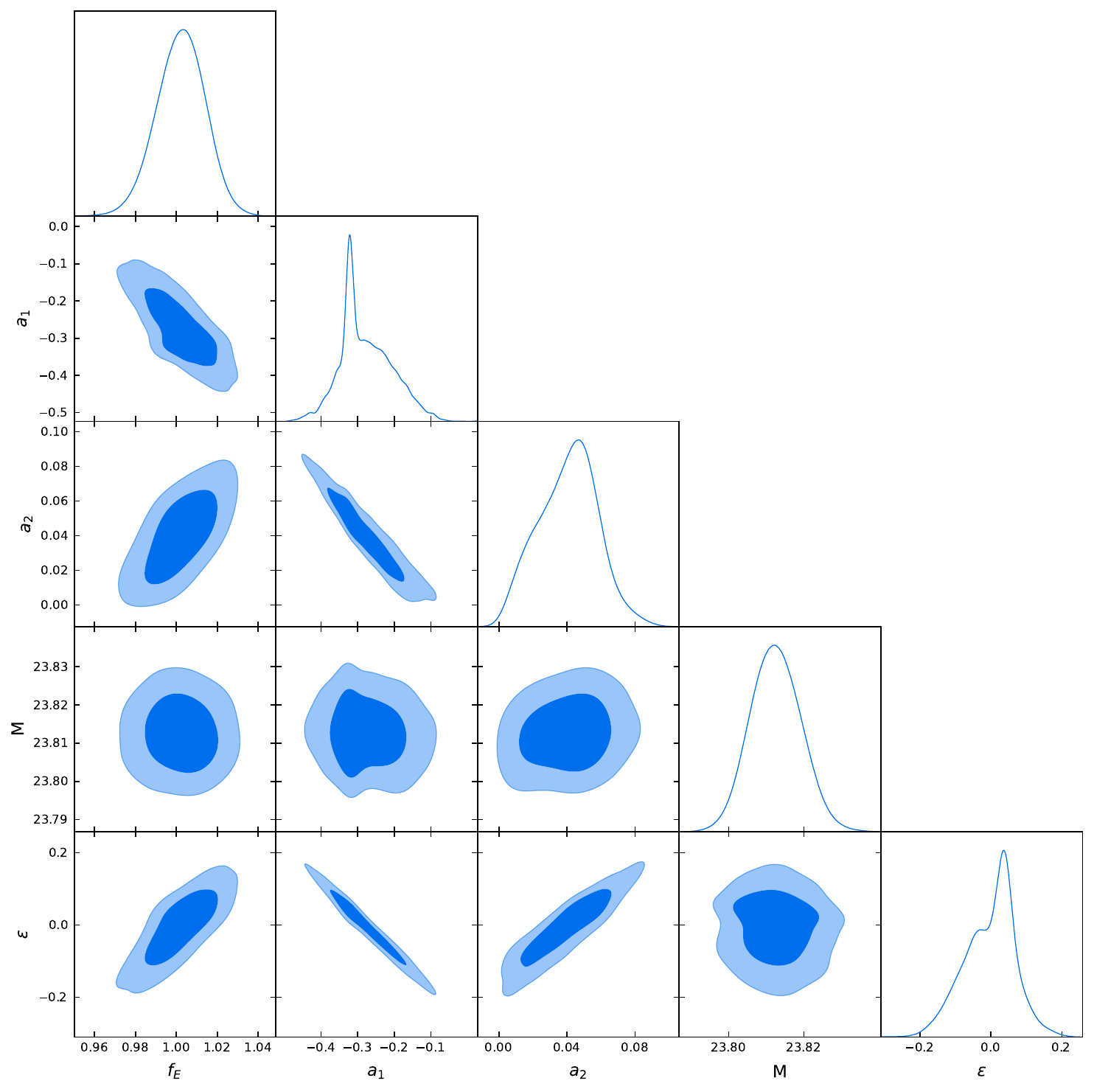}
 \caption{ \small{ Posterior probability distribution functions and 2- D confidence contours of the parameters using 102 SGL systems and complete cosmology-only Pantheon+ dataset in redshift range $ 0.01 < z \leq 2.26137$.}}
  
\end{figure*}

\begin{figure*}
\centering

  \includegraphics[height=.5\textheight]{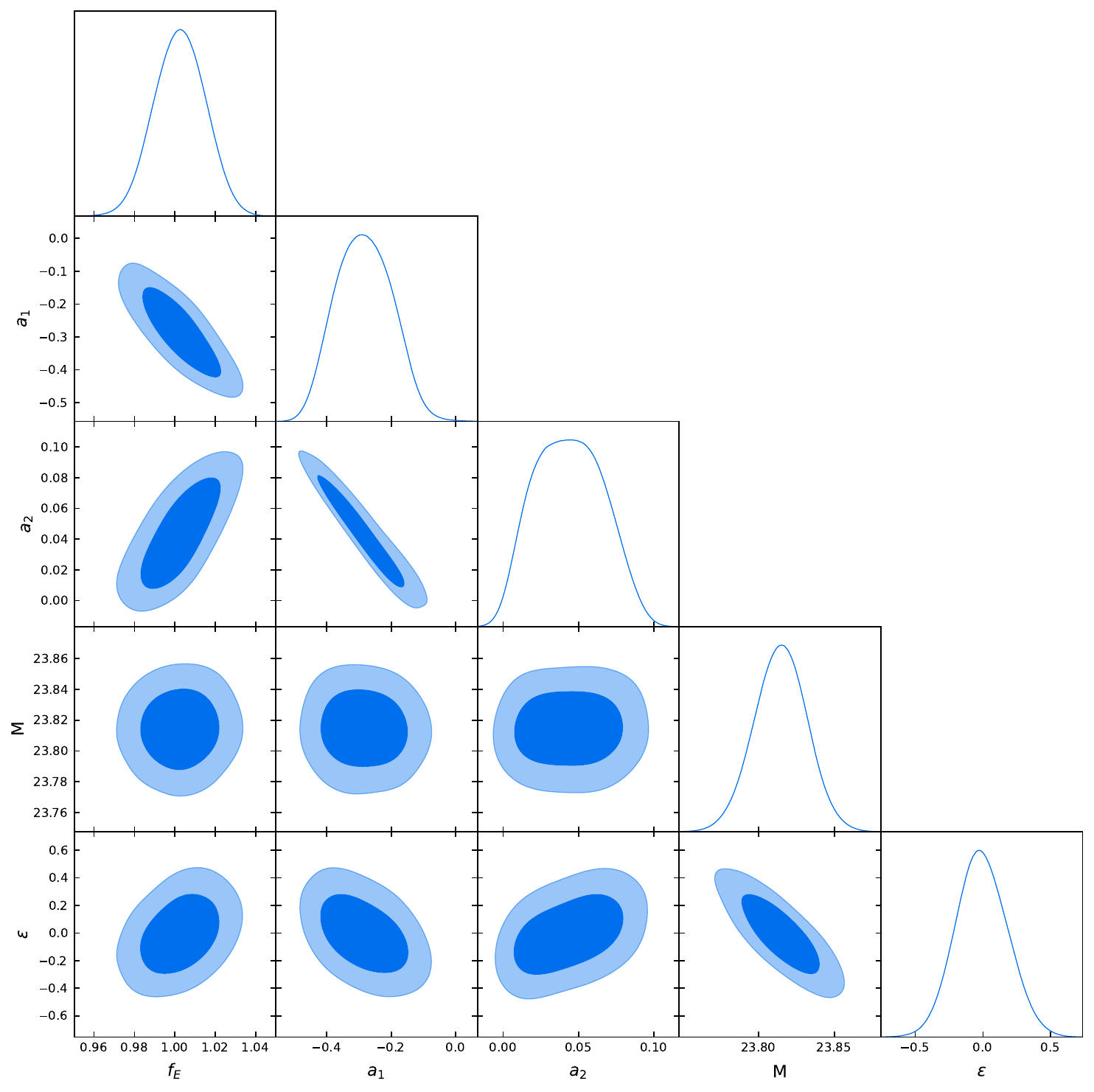}
 \caption{ \small{ Constraint results for the parameters using 102 SGL systems and 630 data points from the Pantheon+ dataset in redshift range $ 0.01 < z \leq 0.1$.}}
  
\end{figure*}

\begin{figure*}
\centering

  \includegraphics[height=.5\textheight]{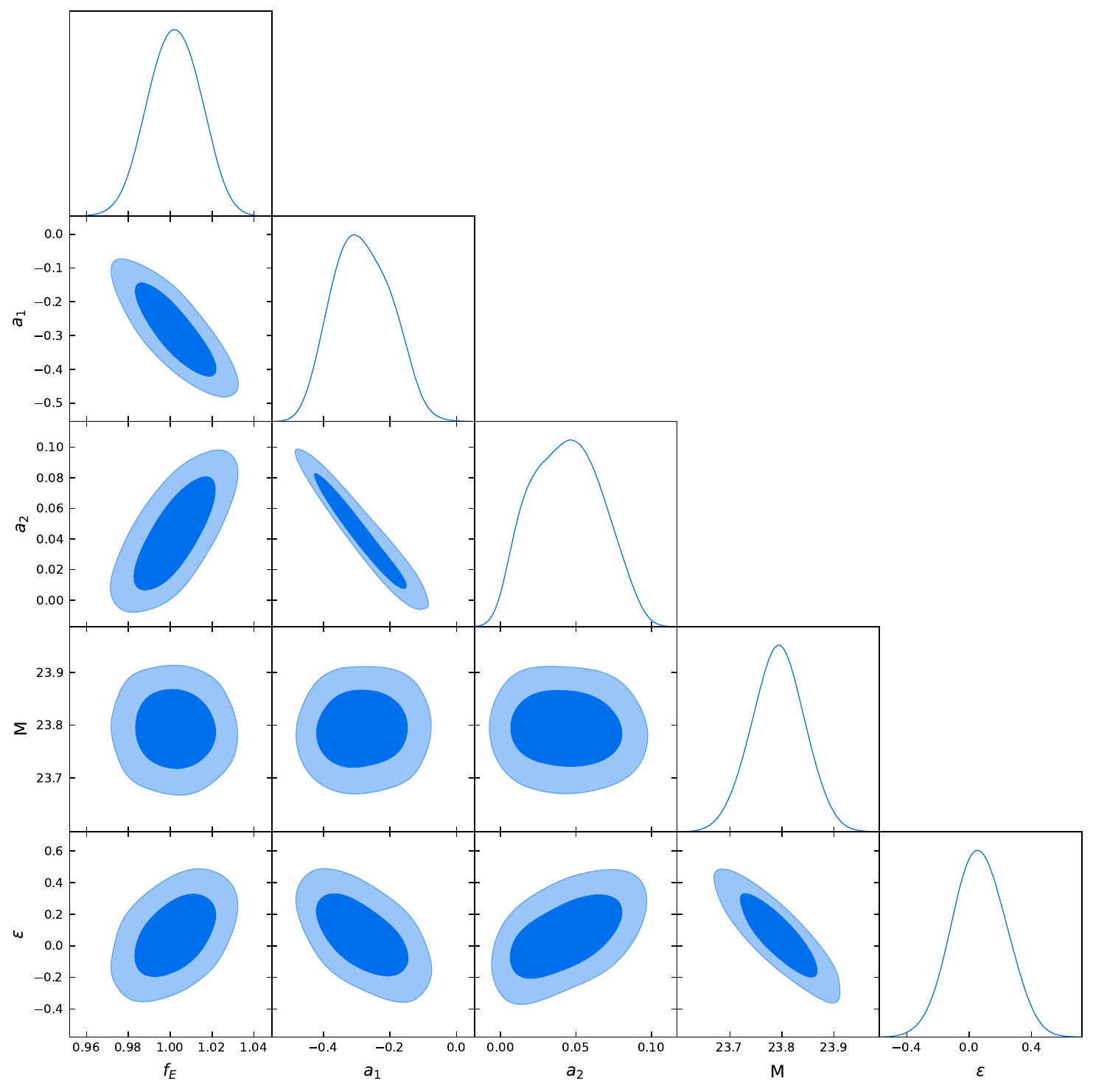}
 \caption{ \small{ Constraint results for the parameters using 102 SGL systems and 207 data points from the Pantheon+ dataset in redshift range $ 0.1 \leq z \leq 0.2$.}}
  
\end{figure*}

\begin{figure*}
\centering

  \includegraphics[height=.5\textheight]{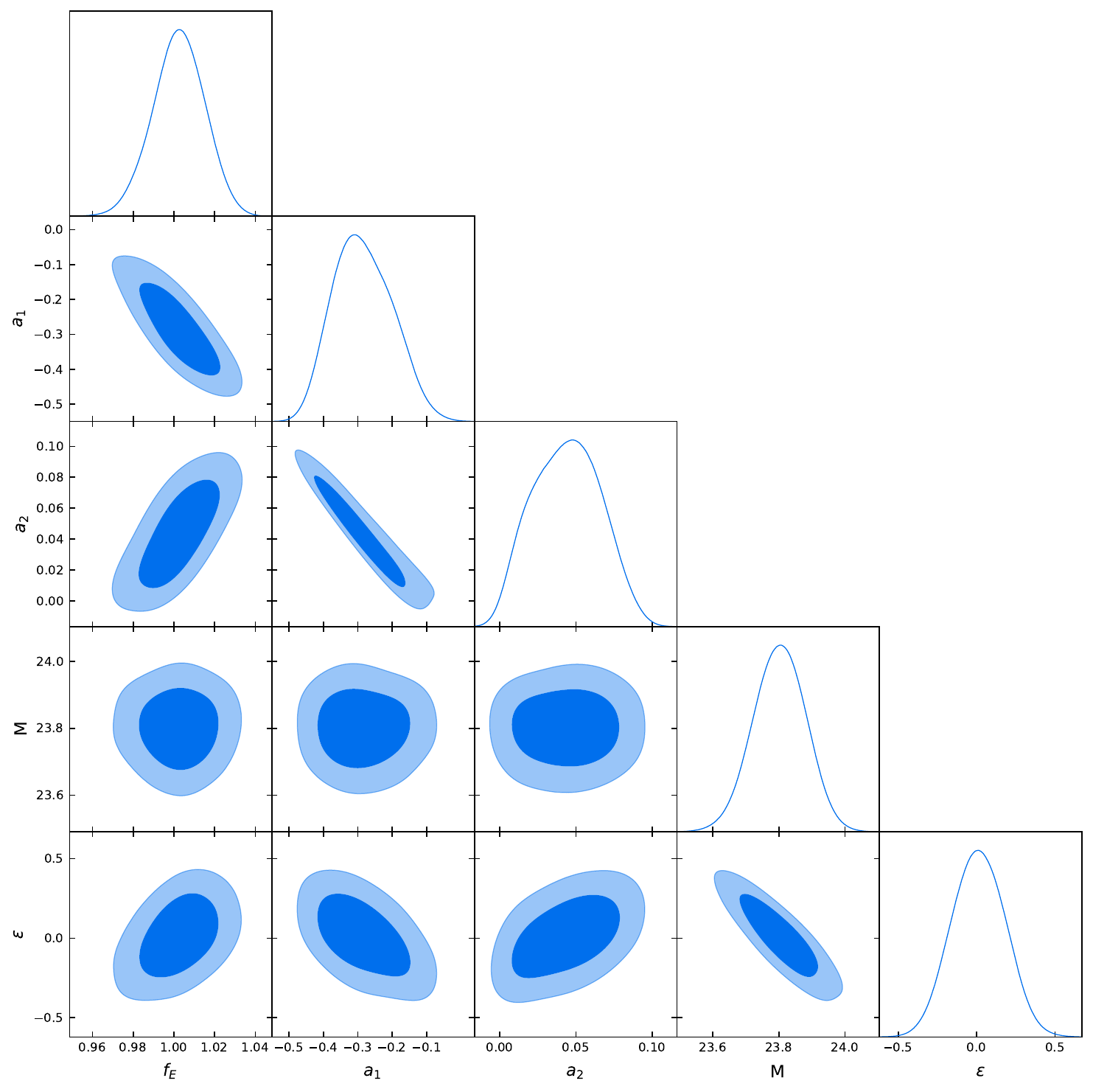}
 \caption{ \small{ Constraint results for the parameters using 102 SGL systems and 259 data points from the Pantheon+ dataset in redshift range $ 0.2 \leq z \leq 0.3 $.}}
  
\end{figure*}

\begin{figure*}
\centering

  \includegraphics[height=.5\textheight]{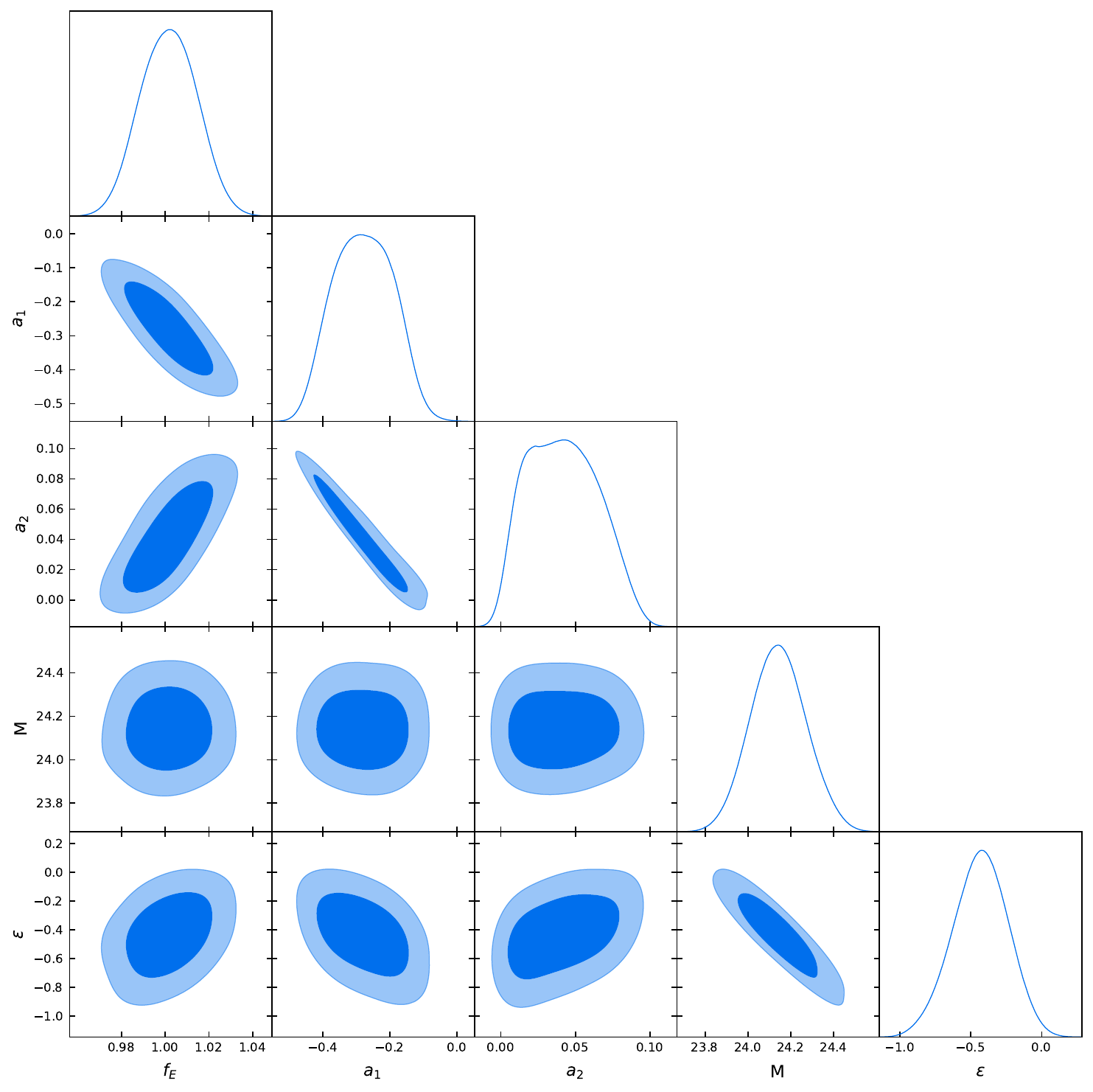}
 \caption{ \small{ Constraint results for the parameters using 102 SGL systems and 186 data points from the Pantheon+ dataset in redshift range $ 0.3 \leq z \leq 0.4$.}}
  
\end{figure*}

\begin{figure*}
\centering

  \includegraphics[height=.5\textheight]{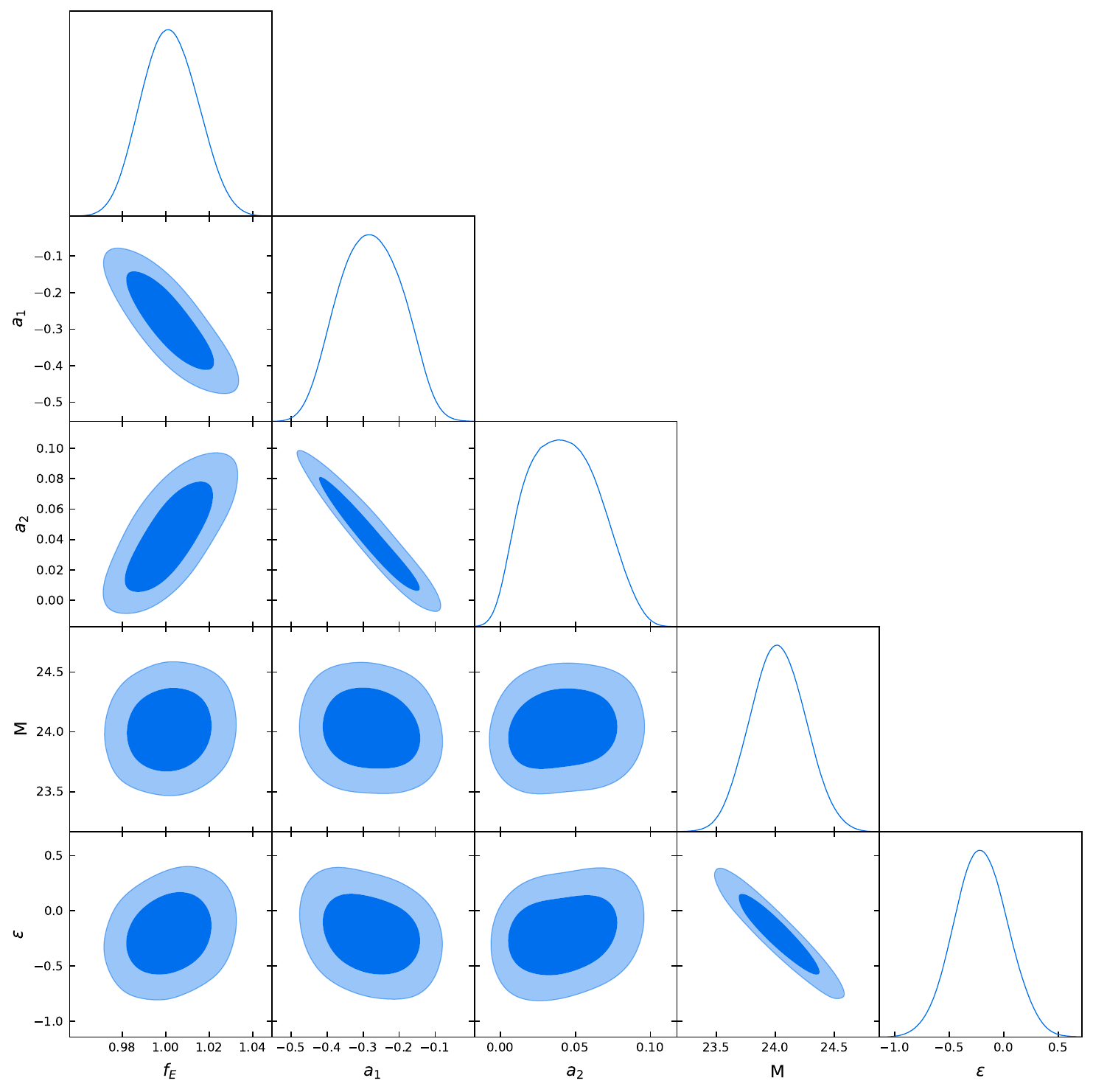}
 \caption{ \small{ Constraint results for the parameters using 102 SGL systems and 98 data points from the Pantheon+ dataset in redshift range $ 0.4 \leq z \leq 0.5$.}}
  
\end{figure*}

\begin{figure*}
\centering

  \includegraphics[height=.5\textheight]{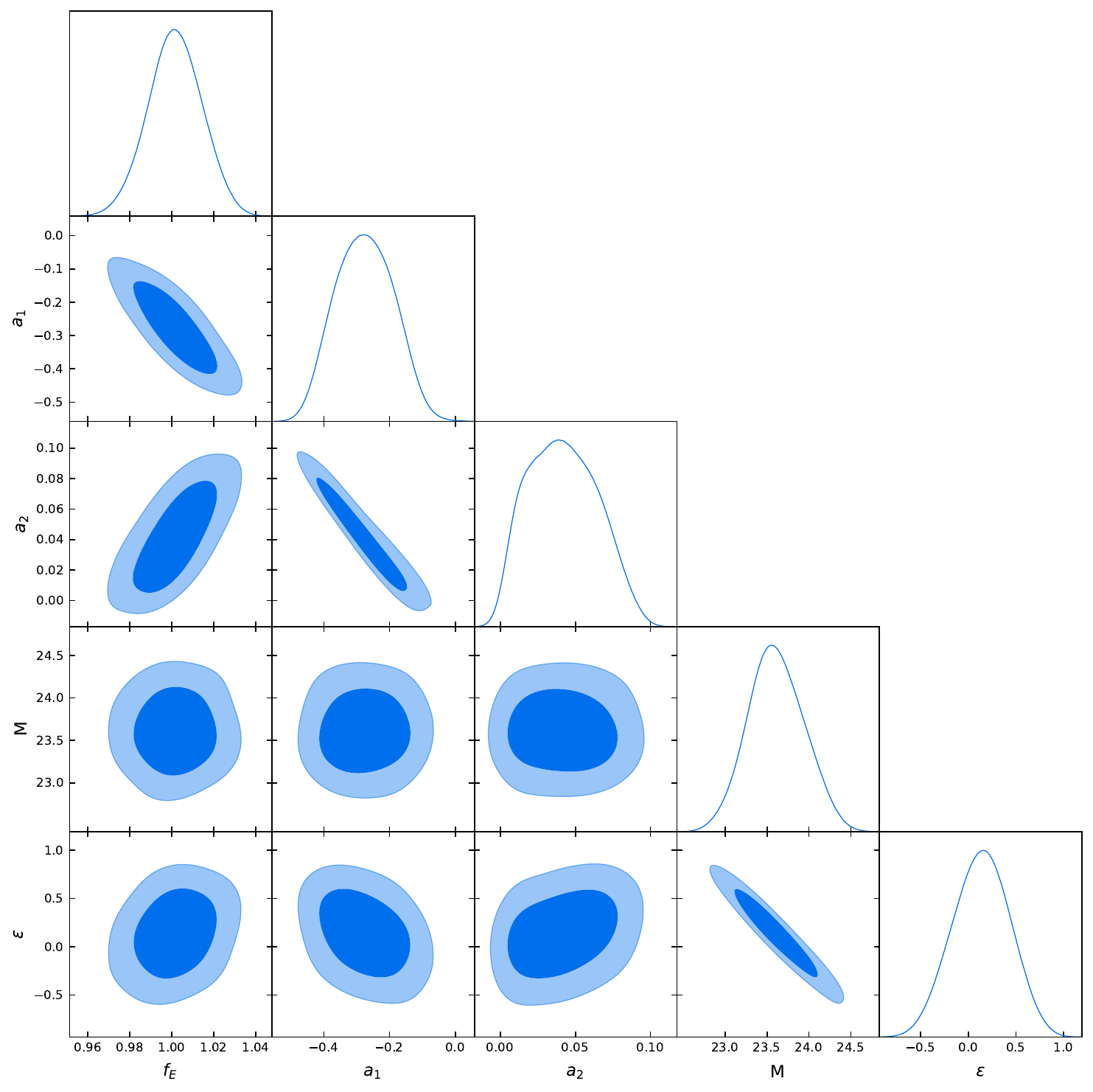}
 \caption{ \small{ Constraint results for the parameters using 102 SGL systems and 81 data points from the Pantheon+ dataset in redshift range $ 0.5 \leq z \leq 0.6$.}}
  
\end{figure*}

\begin{figure*}
\centering

  \includegraphics[height=.5\textheight]{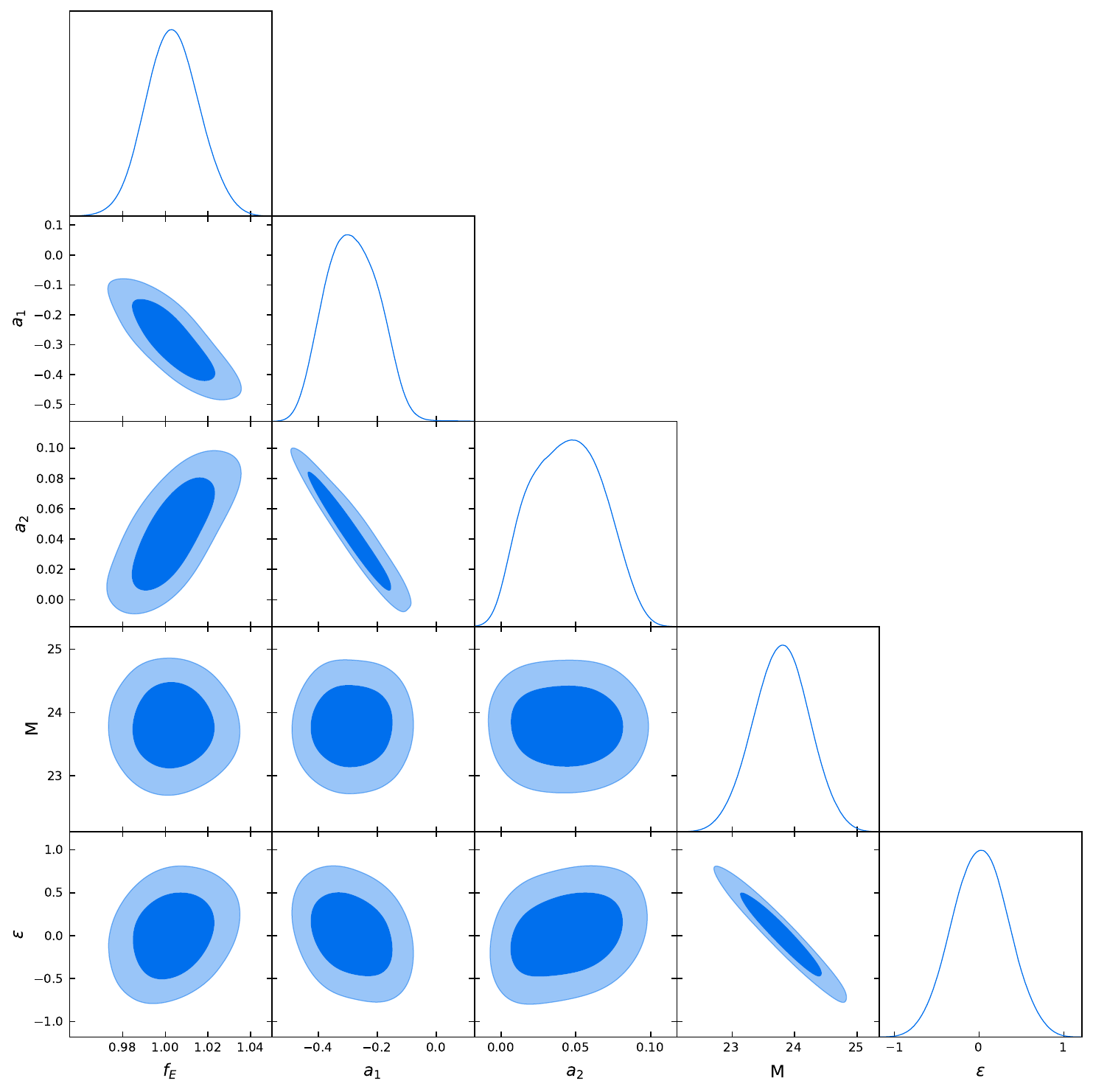}
 \caption{ \small{ Constraint results for the parameters using 102 SGL systems and 54 data points from the Pantheon+ dataset in redshift range $ 0.6 \leq z \leq 0.7$.}}
  
\end{figure*}

\begin{figure*}
\centering

  \includegraphics[height=.5\textheight]{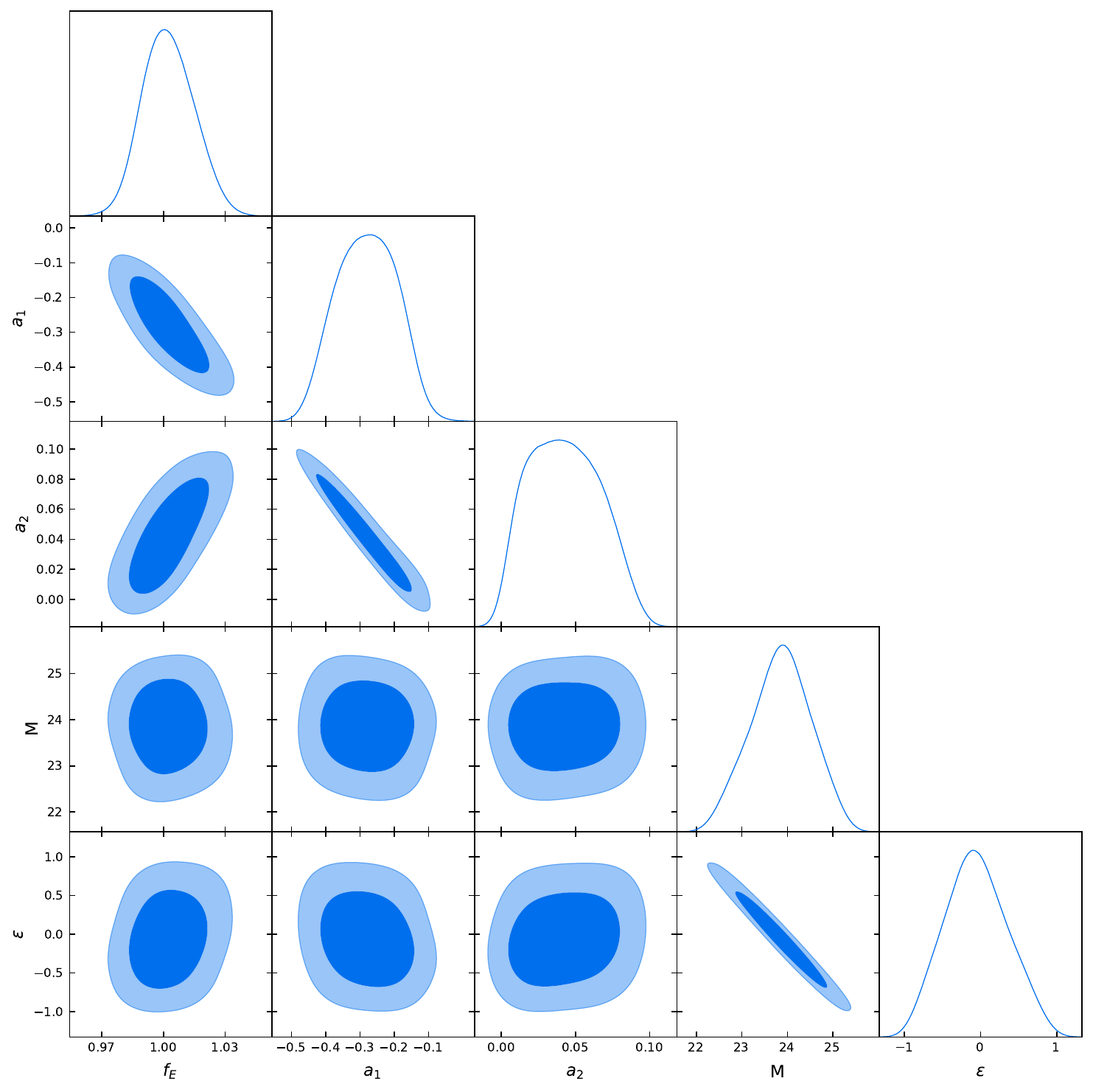}
 \caption{ \small{ Constraint results for the parameters using 102 SGL systems and 45 data points from the Pantheon+ dataset in redshift range $ 0.7 \leq z \leq 0.8$.}}
  
\end{figure*}

\begin{figure*}
\centering

  \includegraphics[height=.5\textheight]{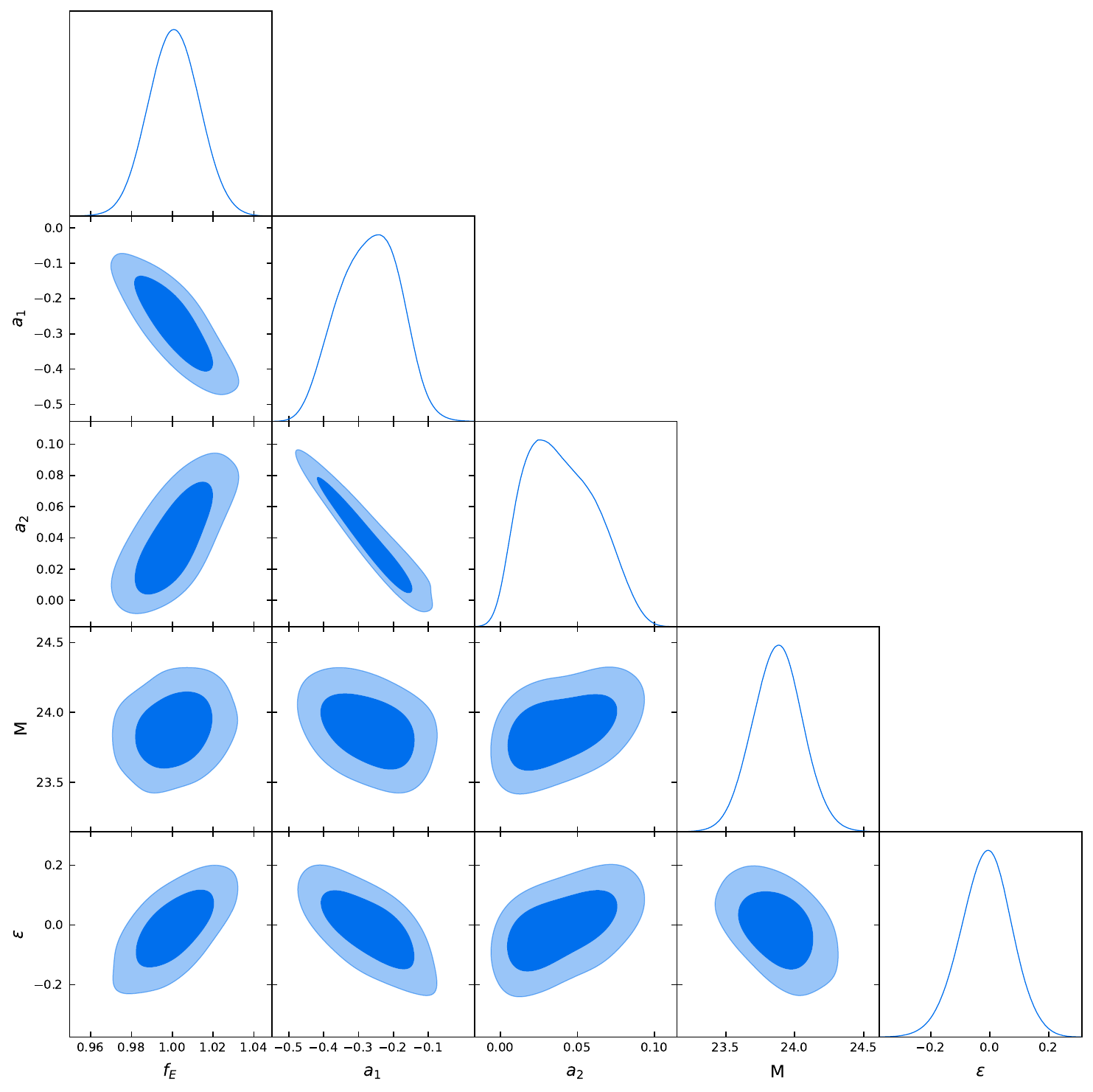}
 \caption{ \small{ Constraint results for the parameters using 102 SGL systems and 30 data points from the Pantheon+ dataset with $  z > 0.8$.}}
  
\end{figure*}

\vspace{-2em}
  
\section*{Acknowledgments} We thank the anonymous referee for constructive and useful comments that helped to improve the manuscript.
\vspace{-1em}
\section*{ Declaration of competing interest} 
The authors declare that they have no known competing financial interests that could have influenced the work reported in this paper.
\vspace{-1em}

\section*{ORCID iD}

Savita Gahlaut    {https://orcid.org/0009-0000-1247-675x} \\
Meetu Luthra      {https://orcid.org/0000-0002-0429-1533} 

\vspace{-1em}




\end{document}